\documentclass[12pt]{article}
\usepackage{amsmath}
\usepackage{amsfonts}
\usepackage{amssymb}
\usepackage{srcltx}

\numberwithin{equation}{section}

\def\rf#1{(\ref{eq:#1})}
\def\lab#1{\label{eq:#1}}
\def\nonu{\nonumber}
\makeatletter
\newcommand{\rd}{\@ifnextchar^{\DIfF}{\DIfF^{}}}
\def\DIfF^#1{%
   \mathop{\mathrm{\mathstrut d}}%
   \nolimits^{#1}\gobblespace}
\def\gobblespace{\futurelet\diffarg\opspace}
\def\opspace{%
   \let\DiffSpace\!%
   \ifx\diffarg(%
   \let\DiffSpace\relax
   \else
   \ifx\diffarg[%
   \let\DiffSpace\relax
   \else
   \ifx\diffarg\{%
   \let\DiffSpace\relax
   \fi\fi\fi\DiffSpace}

\providecommand*{\dder}[3][]{%
\frac{\rd^{#1}#2}{\rd #3^{#1}}}
\providecommand*{\pder}[3][]{%
\frac{\partial^{#1}#2}{\partial #3^{#1}}}

\begin{document}
\vspace*{-1cm}
\noindent
October, 2010 \\
\\
\vskip .3in

\begin{center}

{\large\bf Integrable Origins of Higher Order Painlev\'e Equations}
\end{center}
\normalsize
\vskip .4in

\begin{center}
H. Aratyn

\par \vskip .1in \noindent
Department of Physics \\
University of Illinois at Chicago\\
845 W. Taylor St.\\
Chicago, Illinois 60607-7059\\
\par \vskip .3in

\end{center}

\begin{center}
J.F. Gomes and A.H. Zimerman

\par \vskip .1in \noindent
Instituto de F\'{\i}sica Te\'{o}rica-UNESP\\
Rua Dr Bento Teobaldo Ferraz 271, Bloco II,\\
01140-070 S\~{a}o Paulo, Brazil
\par \vskip .3in

\end{center}
\vskip .2in \noindent
\begin{center}
{\large {\bf ABSTRACT}}\\
\end{center}
\par \vskip .15in \noindent

Higher order Painlev\'e equations invariant under extended affine Weyl groups
$A^{(1)}_n$ are obtained through self-similarity limit
of a class of pseudo-differential Lax hierarchies with 
symmetry inherited from the underlying generalized Volterra lattice
structure.

\newpage
\section{Introduction}
This paper investigates a self-similarity limit of
a special class of  pseudo-differential Lax hierarchies of  the
constrained KP hierarchy with symmetry structure defined by B\"acklund 
transformations
induced by a discrete structure of Volterra type lattice.
The underlying integrable hierarchy is realized in terms of $2 M$ Lax
coefficients $e_i, c_i, \; i=1,{\ldots} ,M$ forming $M$ ``Darboux-Poisson''canonical
pairs  with respect to the second 
Gelfand-Dickey bracket of the underlying constrained 
Kadomtsev-Petviashvili (KP) hierarchy.

It is shown that in the self-similarity limit the second $t_2$-flow 
equations of that hierarchy
transform to the higher order Painlev\'e equations :
\begin{equation}
f_i^{\prime} = f_i \left(f_{i+1}- f_{i+2} + f_{i+3} -
f_{i+4}+{\cdots}-f_{i-1}  \right) + \alpha_i, \quad i=0,1,\ldots,2M 
\lab{hpeqs}
\end{equation}
under a change of variables from
$e_i, c_i, \; i=1,{\ldots} ,M$ to $f_i,\; i=1,{\ldots} ,2 M$,
which is described in subsection \ref{subsect:generalconstruct}.
Equation \rf{hpeqs} introduced $f_0=-\sum_{i=1}^{2M} f_i -2 x$ and 
constants $\alpha_i$ satisfying
$\sum_{i=0}^{2M} \alpha_i =-2$. 
The system satisfies periodicity conditions
$f_{2M+i}=f_{i-1} ,\alpha_{2M+i}=\alpha_{i-1},\; 
i=0,1,2,{\ldots} ,2M$.
These equations are invariant under B\"acklund transformations forming 
the extended affine Weyl group $A^{(1)}_{2M}$.
The extended affine Weyl group $A^{(1)}_{n}$ is generated 
by $n+1$ transformations $s_0, s_1, {\ldots} , s_n$
in addition to cyclic permutation $\pi$
which together satisfy relations
\[ \begin{split}
s_i s_{j} s_i &= s_{j} s_i s_{j}\;\;(j=i \pm 1)
, \qquad s_i s_{j}= s_{j} s_i \;\;(j\ne i \pm 1)\\
\pi s_i &= s_{i+1} \pi, \qquad \pi^{n+1} = 1,\qquad s_i^2=1  \, .
\end{split}
\]
The symmetric Painlev\'e equations with their extended affine Weyl 
symmetry group $A^{(1)}_{n}$
first appeared in Adler's paper \cite{Adler93}
in the setting of periodic dressing chains and later 
were discussed in great details by Noumi and Yamada \cite{NoumiYamada98a,
NoumiYamada98b} (see also \cite{noumibk}).

Imposing second-class constraints on  the
second KP Poisson bracket structure via Dirac scheme reduces a number of  $2 M$ Lax
coefficients to $2M-1$ (or in general $2M-k$) coefficients and via
self-similarity reduction reproduces Painlev\'e equations with
the extended affine Weyl symmetry $A^{(1)}_{2M-1}$. Here we just
present results of the special case of the extended affine 
Weyl symmetry $A^{(1)}_{3}$ and the corresponding 
Painlev\'e V equation.

In section \ref{section:integrable} we present the underlying
integrable Lax hierarchy focusing on the second flow equations
and B\"acklund transformation keeping the Lax equations invariant.
In Section \ref{section:selfsimilar} the self-similarity limit
is taken and Hamiltonians governing the $t_2$ flow equations in 
this limit are derived.
Next, in Section \ref{section:connection} the Hamiltonians
found in Section \ref{section:selfsimilar} are shown to reproduce
Hamiltonian structure of the higher Painlev\'e equations
invariant under the extended affine Weyl symmetry $A^{(1)}_{2M}$
when expressed in terms of canonical variables.
This is illustrated for special cases of $M=1,2,3$ for which the 
generators of the extended affine Weyl symmetry group are derived from
the B\"acklund transformation of Section \ref{section:integrable}.
The Dirac reduction scheme when applied on the original integrable 
KP hierarchy allows reducion of the model with
$A^{(1)}_{2M}$ symmetry down to a model characterized
by $A^{(1)}_{2M-1}$ symmetry.
This is illustrated in Section \ref{section:reduction} for $M=2$ when 
the reduced model is nothing but the  Painlev\'e V system.
Concluding comments are given in Section \ref{section:outlook}.

\section{The Integrable Hierarchy, its Second Flow and Generalized 
Volterra Symmetry Structure}
\label{section:integrable}
\subsection{A ``half-integer'' Lattice}
It is well-known that symmetry of many continuum KP-type hierarchies 
are governed by discrete lattice-like structures. A standard example is provided
by the AKNS hierarchy and the Toda lattice structure of its B\"acklund
transformations leading to Hirota type equations for the Toda chain of
tau-functions \cite{Aratyn:1993zi}.
There also exists the  so-called two-boson formulation of the AKNS
hierarchy which is invariant under symmetry transformations on 
a  ``half-integer'' lattice which generalizes Toda lattice \cite{Aratyn:1993zi}.
We now present a general ``half-integer'' lattice (or 
the generalized Volterra lattice) following closely reference \cite{Aratyn:1994vc}.
The foundation of this formalism rests on two spectral equations:
\begin{align}
\lambda^{1/2}\; {\widetilde \Psi}_{n+\frac12} 
&=  \Psi_{n+1} + {\cal A}^{(0)}_{n+1} \Psi_n +
\sum_{p=1}^M {\cal A}^{(p)}_{n-p+1} \Psi_{n -p} \lab{sqraa} \\
\lambda^{1/2} \; \Psi_n &= {\widetilde \Psi}_{n+\frac12} + 
{\cal B}^{(0)}_n {\widetilde \Psi}_{n-\frac12}
\lab{sqrbb}
\end{align}
and ``time'' evolution equations:
\begin{equation}
{\widetilde \Psi}_{n+\frac12} = \left( \partial - {\cal B}^{(0)}_n - {\cal A}^{(0)}_{n} \right){\widetilde \Psi}_{n-\frac12}
\quad;\quad
\Psi_{n+1} = \left( \partial - {\cal B}^{(0)}_n - {\cal A}^{(0)}_{n+1} \right) \Psi_{n}
\lab{evolution}
\end{equation}
which both involve objects labeled by integers and half-integers.
After removing the term $ \sum_{p=1}^M {\cal A}^{(p)}_{n-p+1} \Psi_{n
-p}$ from equation \rf{sqraa} the above system yields the Volterra
chain equations. For that reason we will refer to equations \rf{sqraa}--\rf{evolution}
as a generalized Volterra system.
As shown in \cite{Aratyn:1994vc}, upon
eliminating the half-integer modes, the generalized Volterra system
\rf{sqraa}--\rf{evolution} reduces to the Toda lattice
equations. 
{}From \rf{sqraa}--\rf{evolution} we find:
\begin{equation}
\begin{split}
\lambda^{1/2}\; {\widetilde \Psi}_{n+\frac12}&= \biggl( \partial - {\cal B}^{(0)}_n +
   \sum_{p=1}^M
{\cal A}^{(p)}_{n-p+1} ( \partial - {\cal B}^{(0)}_{n-p} - {\cal A}^{(0)}_{n-p+1} )^{-1}
\cdots \biggr. \\
&\biggl. \cdots ( \partial - {\cal B}^{(0)}_{n-1} - {\cal A}^{(0)}_{n} )^{-1}
\biggr) \Psi_n
\lab{speca}
\end{split}
\end{equation}
and
\begin{equation}
\lambda^{1/2}\; \Psi_{n}= \left( \partial - {\cal A}^{(0)}_n \right) 
{\widetilde \Psi}_{n-\frac12}
\lab{specb}
\end{equation}
Eliminating half-integer modes from the last two relations 
yields a spectral equation of  a form 
\begin{equation}
 \lambda \Psi_n =L^{(M+1)}_n
\Psi_{n}
\lab{psilm1}
\end{equation}
with Lax operator $L^{(M+1)}_n$
given by recurrence relation:
\begin{equation}
L^{(M+1)}_n = e^{\int {\cal B}^{(0)}_{n-1}} \left( \partial - {\cal A}^{(0)}_n +{\cal B}^{(0)}_{n-1}\right)
L^{(M)}_n \left( \partial - {\cal A}^{(0)}_n \right)^{-1} e^{-\int {\cal B}^{(0)}_{n-1}}
\lab{recu}
\end{equation}
where
\begin{equation}
L^{(M)}_n=  \partial  + \sum_{p=1}^M
{\cal A}^{(p)}_{n-p} ( \partial + {\cal B}^{(0)}_{n-1}- {\cal B}^{(0)}_{n-p-1} -
{\cal A}^{(0)}_{n-p} )^{-1}
\cdots ( \partial + {\cal B}^{(0)}_{n-1}-{\cal B}^{(0)}_{n-2} - {\cal A}^{(0)}_{n-1} )^{-1}
\lab{laxMn}
\end{equation}
Using equation \rf{specb} it is easy to shift the spectral
equation \rf{psilm1} to the half-integer lattice:
\begin{equation}
\lambda {\widetilde \Psi}_{n-\frac12}
= {\widetilde L}^{(M+1)}_n {\widetilde \Psi}_{n-\frac12}, \quad \;\;
{\widetilde L}^{(M+1)}_n = \left( \partial - {\cal A}^{(0)}_n \right)^{-1}
L^{(M+1)}_n \left( \partial - {\cal A}^{(0)}_n \right) 
\lab{backpsilm1}
\end{equation}
The similarity transformation responsible for 
transformation from integer to half-integer lattice will
be shown below to play a central role as a B\"acklund transformation
of the higher order Painlev\'e equations.

\subsection{Basic facts about $2M$-bose constrained KP hierarchy}
The recurrence relation \rf{recu} is realized by 
the $2M$-bose constrained KP hierarchy with
Lax operators $L_{M}$, $M=1,2,\ldots$ :
\begin{align}
L_{M} &=  \left( \partial - e_{M} \right) \prod_{k=M-1}^1 \left( \partial - e_k -
\sum_{l=k+1}^{M} c_l \right) \left( \partial - \sum_{l=1}^{M} c_l \right) \nonu\\
&\times \prod_{k=1}^{M}
\left( \partial - e_k - \sum_{l=k}^{M} c_l \right)^{-1}  \lab{1c}
\end{align}
given here in terms of the ``Darboux-Poisson''canonical pairs $(c_k,e_k)_{k=1}^{M}$.
Recall that  the KP hierarchy is endowed with bi-Hamiltonian
Poisson bracket structures resulting from the two compatible Hamiltonian
structures on the algebra of pseudo-differential operators.
Remarkably, for the above Lax hierarchy the second bracket of hierarchy is 
realized in terms  of $(c_k,e_k)_{k=1}^{M}$
as a Heisenberg Poisson bracket algebra:
\begin{equation}
\left\{ e_i (x), c_j (y)\right\}_2 = - \delta_{ij} \delta_x (x-y), \;\;
i,j =1,2, \ldots, M
\lab{2bracket}
\end{equation}
The Lax operator \rf{1c} realizes the recursive relation 
\rf{recu} rewritten in this context as follows:
\begin{equation}
L_{M} = e^{\int c_{M}} \left( \partial + c_{M} - e_{M} \right) L_{M-1}
\left( \partial - e_{M} \right)^{-1} e^{-\int c_{M}} 
, \quad L_0 \equiv \partial 
\lab{recuce}
\end{equation}
for $M=1,2, \ldots $.
The corresponding second flow equations can be obtained from the second
bracket structure as follows 
\begin{equation}
\pder{f}{t_2}= \left\{ f , H_2 \right\}_2\, , 
\lab{ft2}
\end{equation}
where the Hamiltonian $H_2$ 
is an integral of the coefficient $u_1 (M)$ appearing in front of 
$\partial^{-2}$ in the Lax operator \rf{recuce} when cast in a conventional KP form :
\[
L_{M}  =\partial+u_0 (M) \partial^{-1}+u_1 (M) \partial^{-2} +{\ldots} 
\]
As a consequence of equation \rf{recuce}
we obtain the recursive relations for the coefficients :
\[
\begin{split}
u_0 (M) &= u_0 (M-1) + \left(e^{\prime}_{M} + e_{M} c_{M} \right)\\
u_1 (M) &= u_1 (M-1) +u_0^{\prime}  (M-1)+ 2 u_0 (M-1) c_{M}
+ \left(e^{\prime}_{M} + e_{M} c_{M} \right) 
\left(e_{M}+ c_{M} \right) \, .
\end{split}
\]
with solutions
\[
\begin{split}
u_0 (M) &= \sum_{i=1}^{M}
\left(e^{\prime}_{i} + e_{i} c_{i} \right) \\
u_1 (M) 
&= \sum_{i=1}^{M-1} (M-i) \left(e^{\prime}_{i} + e_{i} c_{i} \right)^{\prime} + 
2 \sum_{i=1}^{M-1} u_0  (i) c_{i+1} +\sum_{i=1}^{M}
\left(e^{\prime}_{i} + e_{i} c_{i} \right) \left( e_{i}+ c_{i} \right) \\
\end{split}
\]
The Darboux-B\"acklund transformation of the Lax operator $L_M$ 
defined in equation \rf{1c} takes a form
\[
L_M \to  \left( \partial - e_{M} \right)^{-1} L_M \left( \partial - e_{M} \right)
\lab{simM}
\]
and since  $e_M \sim {\cal A}^{(0)}_n $ we see from equation \rf{specb}
that it represents transformation on the Volterra lattice from
integer modes to half-integer modes.
In terms of coefficients this results
for coefficients with highest indices  in :
\begin{equation}
\begin{split}
g( e_M)&=e_{M-1} + c_M, \quad g( c_M)=-e_{M-1} + e_M-\frac{c_M^{\prime}}{c_M}\\
g( e_{M-1})&=e_{M-2}+e_{M-1}-e_M +c_M+c_{M-1} + \frac{c_M^{\prime}}{c_M} \\
g( c_{M-1})&=-e_{M-2} + e_{M-1} -\frac{\left(   -e_{M-1} + e_M-c_M-c_{M-1} -\frac{c_M^{\prime}}{c_M} \right)^{\prime}}{\left(-e_{M-1} + e_M-c_M-c_{M-1} -\frac{c_M^{\prime}}{c_M}\right)}
\end{split}
\lab{gemcm}
\end{equation}
in addition to
\begin{equation}
g \left( e_k+\sum_{l=k+1}^M c_l \right)=e_{k-1}+\sum_{l=k}^M c_l, \;\; 2 \le k \le M,\;\;
g \left( e_1+\sum_{l=2}^M c_l \right)=\sum_{l=1}^M c_l\, .
\lab{gaddition}
\end{equation}
For a special example of the so-called two Bose system 
with $M=1$ (which below will
be shown to correspond to the symmetric Painlev\'e IV
equations)
the Lax operator is:
\[
L_1 = (\partial-e_1)(\partial-c_1)(\partial-e_1-c_1)^{-1}
\]
Such a  Lax operator possesses a Darboux-B\"acklund symmetry:
\[
L_1 \to (\partial-e_1)^{-1} L_1 (\partial-e_1)= (\partial-c_1)(\partial-e_1+c_{1\,x}/c_1)(\partial-e_1-c_1+c_{1\,x}/c_1)^{-1}
\]
which keeps its form  unchanged and transforms $e_1,c_1$ as 
follows :
\begin{equation}
g \left(e_1 \right) = c_1 , \;\; g  \left(c_1 \right)=e_1- c_{1\,x}/c_1, 
\lab{g2bose}
\end{equation}
\section{Hamiltonians and $\mathbf{t_2}$ Flow Equations in 
the Self-similarity  Limit of the $2M$-bose constrained KP hierarchy}
\label{section:selfsimilar}
Second flow equation \rf{ft2} results in the following 
expressions for the Lax coefficients:
\begin{equation}
\begin{split}
\pder{c_{j}}{t_2}&= \dder{}{x}\left( c^{\prime}_{j}-
c^{2}_{j}-2 e_{j} c_{j}+2 \sum_{i=j+1}^{M} c^{\prime}_{i}
-2 \sum_{i=j}^{M-1} c_{i}c_{i+1}
\right), \;\;\; j =1, {\ldots} , M \\
\pder{e_{j}}{t_2}&= \dder{}{x}\left( -e^{\prime}_{j}-
e^{2}_{j}-2 e_{j} c_{j}-2 u_0 (j-1)-2 e_{j}\sum_{i=j+1}^{M} c_{i}
\right), \;\;\; j =1, {\ldots} , M
\end{split}
\lab{t2flow}
\end{equation}
Effectively, the action of the self-similarity reduction replaces 
$\partial f /\partial t_2$ with $-(x f)_x/2$.
Integrating all equations obtained through taking the self-similarity limit  we find
\begin{equation}
\begin{split}
e^{\prime}_j+2 \sum_{i=1}^{j-1} e^{\prime}_i&=
2 x e_{j} -e^2_j -2 e_j \left( \sum_{i=j}^{M-1} c_{i+1}  \right)
-2 \sum_{i=1}^j e_ic_i +{\bar \kappa}_j\\
c^{\prime}_j+2 \sum_{i=j+1}^{M} c^{\prime}_i&=
-2 x c_{j}+ c^{2}_{j} +2 c_j \sum_{i=j}^{M-1}c_{i+1} + 2c_j e_j
+{\kappa}_j 
\lab{ejcjeqs}
\end{split}
\end{equation}
for $j=1, {\ldots} , M$ and with integration constants
$\kappa_j, {\bar \kappa}_j$.
The above equations are Hamiltonian in a sense that :
\begin{equation}
e^{\prime}_j+2 \sum_{i=1}^{j-1} e^{\prime}_i = \pder{{\cal H}_M}{c_j},
\;\;\quad
c^{\prime}_j+2 \sum_{i=j+1}^{M} c^{\prime}_i= - \pder{{\cal H}_M}{e_j}
\lab{ejcjHeqs}
\end{equation}
with
\begin{equation}
{\cal H}_M 
= - \sum_{j=1}^M e_j c_j  \left( e_j+c_j -2 x\right)
-2 \sum_{1\le j <i \le M} e_j c_j c_i +\sum_{j=1}^M {\bar \kappa}_j c_j - \sum_{j=1}^M {\kappa}_j e_j
\lab{calhm}
\end{equation}
Note, that the Hamiltonian ${\cal H}_M$ defined in \rf{calhm}
satisfy 
\begin{equation}
\dder{{\cal H}_M}{x}=\sum_{j=1}^{M}\left( \pder{{\cal H}_M}{e_j} e_j^\prime+
\pder{{\cal H}_M}{c_j} c_j^\prime  \right)+\pder{{\cal H}_M}{x}= \pder{{\cal
H}_M}{x}=
2  \sum_{j=1}^M e_j c_j
\lab{hmlemma}
\end{equation}
as follows from the fact that the first two terms on the right hand
side cancel.

It follows that equations \rf{ejcjeqs}  can be rewritten as  
\begin{align}
e^{\prime}_j&=2 x e_j + 4 x \sum_{k=1}^{j-1} (-1)^{j-k} e_k
- e_j \left( e_j+2  \sum_{k=j}^{M} c_k\right) \nonumber \\
&+\sum_{k=1}^{j-1} (-1)^{j-k+1} e_k\left(2 e_k+2c_k+
4  \sum_{l=k+1}^{M} c_l \right) +{\bar k}_j \lab{eprj}\\
c^{\prime}_j&=-2 x c_j + 4 x \sum_{k=j+1}^{M} (-1)^{j-k+1} c_k
+ c_j \left( c_j + 2 e_j+2  \sum_{k=j+1}^{M} c_k\right) \nonumber \\
&+\sum_{k=j+1}^{M} (-1)^{j-k} c_k\left(2 c_k+4e_k+
4  \sum_{l=k+1}^{M} c_l \right) +{k}_j \lab{cprj}
\end{align}
for $j=1, {\ldots} , M$ with appropriately redefined constants $k_j,
{\bar k}_j$:
\[
\kappa_j=k_j+2 \sum_{i=j+1}^M k_i, \quad
{\bar \kappa}_j={\bar k}_j+2 \sum_{i=1}^{j-1} {\bar k}_i,\;\;\;\;
j=1,{\ldots} , M
\]
Equations \rf{eprj}-\rf{cprj} can be expressed as 
actions of two vector fields on ${\cal H}_M$ :
\begin{equation}
e^{\prime}_j= E_j \left( {\cal H}_M \right), \qquad
c^{\prime}_j=C_j \left( {\cal H}_M \right)
\lab{ecjhm}
\end{equation}
where vector fields $ E_j , C_j$ are
\begin{equation}
E_j = \sum_{i=1}^{j} (-1)^{j-i}
\left( \pder{}{c_i}- \pder{}{c_{i-1}} \right) ,\quad C_j = - \sum_{i=j}^{M}  (-1)^{i-j} \left( \pder{}{e_i} - \pder{}{e_{i+1}} \right)
\lab{ectflds}
\end{equation}
We now can define the Poison brackets through
\begin{equation}
\left\{ e_j \, ,\, F\right\}= E_j \left( F \right), \qquad
\left\{ c_j \, ,\, F\right\}=
C_j \left( F \right)
\lab{pbec}
\end{equation}
which results in a Poisson bracket structure :
\begin{equation}
\left\{ e_j \, ,\, c_i \right\}= \delta_{j,i}+2 E_{j,i}
\lab{pbarray}
\end{equation}
where $E_{j,i}$ is an element of strictly lower-triangular matrix
and equal to:
\[
E_{j,i}=\begin{cases} (-1)^{j-i} &  j>i, \\
 0 & j \le i \, .
\end{cases} 
\]
The equations of motion \rf{eprj}-\rf{cprj} are reproduced through
\begin{equation}
e_j^{\prime} = \left\{ e_j \, ,\, {\cal H}_M  \right\}, \qquad
c_j^{\prime} = \left\{ c_j \, ,\, {\cal H}_M  \right\}\, .
\lab{pbhamm}
\end{equation}

\section{Connection to Higher Order Painlev\'e Equations}
\label{section:connection}
\subsection{General Construction}
\label{subsect:generalconstruct}
Let $q_i, p_i, \, i=1,{\ldots} ,M$ be canonical coordinates 
satisfying the  canonical brackets
\[
\left\{ q_i\, , \, p_{j} \right\} =  -\delta_{ij}, \;\;\;
\left\{ q_i\, , \, q_{j} \right\} = 0=
\left\{ p_i\, , \, p_{j} \right\}, \;\;\;\; i=1,{\ldots} ,M\, .
\]
Relations 
\begin{equation}
q_i = f_{2i}, \qquad p_i= \sum_{k=1}^i f_{2k-1},\;\;\;\;\;\;\;  i=1,{\ldots} ,M
\lab{pqtof}
\end{equation}
define new variables $f_k, \, k=1, {\ldots} ,2M$ and map
the canonical brackets into the following Poisson brackets :
\[
\left\{ f_i\, , \, f_{i+1} \right\} = 1, \qquad
\left\{ f_i\, , \, f_{i-1} \right\} =  -1, \;\; i=1,{\ldots} ,2M\, .
\]
We now propose a conversion table
mapping $e_i, c_i \; i=1, {\ldots}, M$
into a special set of canonical coordinates as well as  
Painlev\'e variables $f_k, \, k=1, {\ldots} ,2M$ that will satisfy 
the higher order Painlev\'e equations \rf{hpeqs}.

First, we list the result for $e_i \; i=1, {\ldots}, M$:
\begin{equation}
\begin{split}
e_M &= q_M+p_M+2x+\frac{k_M}{p_M-p_{M-1}}= \sum_{i=1}^M f_{2i-1}+f_{2M}
+2x+\frac{k_M}{f_{2M-1}}\\
e_{M-1} &= -p_{M-1}= -\sum_{i=1}^{M-1} f_{2i-1}\\
e_{M-2}&=-q_1-\cdots -q_{M-2}=-f_2-\cdots - f_{2M-4}\\
e_{M-3}&=-p_{M-2}+p_1=-f_3-f_5-\cdots-f_{2M-5}\\
\cdots &= \cdots \\
e_2&=-f_{M-2}-f_{M} =\begin{cases} -q_{M/2}-q_{M/2-1} & \text{$M$ even}\\
-p_{(M+1)/2}+p_{(M-3)/2} & \text{$M$ odd}
\end{cases}\\
e_1&= - f_{M-1}= \begin{cases} -p_{M/2}+p_{M/2-1} & \text{$M$ even}\\
-q_{(M-1)/2} & \text{$M$ odd}
\end{cases}
\end{split}
\lab{eMM}
\end{equation}
and next for $c_i \; i=1, {\ldots}, M$ :
\begin{equation}
\begin{split}
c_M &= -p_M+p_{M-1}=-f_{2M-1},\\
c_{M-1}& =p_M-p_{M-1}-q_M-q_{M-1} =f_{2M-1} -f_{2M-2}-f_{2M}\\
c_{M-2} &= p_{M-2}+q_1+q_2+ \cdots+q_{M-3} +q_{M-2}+2 q_{M-1}+2 q_{M}+2x\\
c_{M-3} &=- p_{M-2}-p_{1}-q_1-q_2 -\cdots-q_{M-4}-
q_{M-3}-2q_{M-2}
\\&-2q_{M-1}-2q_M-2x, \\
c_{M-4} &=p_1+p_{M-3}+q_2+\cdots+q_{M-4}+q_{M-3}+\\&+2q_{M-2}+2q_{M-1}+2q_M+2x\\
\cdots &= \cdots \\
c_{M-2k} &= p_{k-1}+p_{M-k-1}+q_k + \cdots + q_{M-k-1} + 2 q_{M-k}
+2q_{M-k+1}
+\cdots \\
&\cdots+2q_{M-1}+2q_M+2x, \quad\qquad k=1,2,3,{\ldots} \\
c_{M-(2k-1)} &=-\left(p_{k-1}+p_{M-k}+q_{k-1}+\cdots+q_{M-k-1}
+2q_{M-k}+\cdots \right.
\\  &\left.+2q_{M-1}+2q_M+2x\right), \quad\qquad k=2,3,{\ldots} 
\end{split}
\lab{cMM}
\end{equation}
Thus from eq. \rf{cMM} it follows that
\begin{equation}
c_1 = \begin{cases} -\left(p_{M/2-1}+p_{M/2}+q_{M/2-1} +2
q_{M/2}+{\ldots} +
+2q_M+2x\right) & \text{$M$ even}\\
p_{(M-3)/2}+p_{(M-1)/2}+q_{(M-1)/2} & \\
+2q_{(M+1)/2}+{\ldots} 
+2q_M+2x
 & \text{$M$ odd}
\end{cases}
\lab{cMM1}
\end{equation}
The Hamiltonian ${\cal H}_M$ defined in \rf{calhm} 
reads in terms of 
$q_i, p_i, \, i=1,{\ldots} ,M$ defined in equations 
\rf{eMM}-\rf{cMM} as follows
\begin{equation} 
\begin{split}
{\cal H}_M&= \sum_{j=1}^M p_j q_j  \left( p_j+q_j +2 x\right)
+2 \sum_{1\le j <i \le M} p_j q_j q_i\\
&-\sum_{j=1}^M {\alpha}_{2j} p_j + \sum_{j=1}^M q_j \left(\sum_{k=1}^j
\alpha_{2k-1}\right)
\end{split}
\lab{calhmpq}
\end{equation} 
in agreement with reference \cite{sasano}.
The corresponding Hamilton equations:
\[
\begin{split}
p_i^\prime &= \pder{{\cal H}_M}{q_i}= p_i  \left( p_i+2 \sum_{j=i}^M 
q_j +2 x\right) +2 \sum_{j=1}^{i-1} p_jq_j+\sum_{j=1}^i \alpha_{2j-1} \\
q_i^\prime &= -\pder{{\cal H}_M}{p_i}= - q_i  \left( 2p_i+q_i+2
\sum_{j>i} q_j +2 x\right) + \alpha_{2i}
\end{split}
\]
are equivalent to the higher Painlev\'e equations as given in
equation \rf{hpeqs} with identification of variables provided
by relation \rf{pqtof}.

In the following subsections of this section we will
illustrate the above general result for $M=1,2,3$.
\subsection{The case of $M=1$ and Painlev\'e IV Equations}
For $M=1$ the equations \rf{eprj}-\rf{cprj} take  the form of the Levi system :
\begin{equation}
\begin{split}
e_1^\prime&=2xe_1-(e_1+2c_1) e_1+{\bar k}_1\\
c_1^\prime &= -2xc_1+(c_1+2e_1)c_1+k_1\, ,
\end{split}
\lab{levi}
\end{equation}
which  is kept invariant under transformations
\rf{g2bose} when accompanied by transformations 
$g(k_1) = 2-{\bar k}_1, \; g \left(  {\bar k}_1 \right)=-k_1$
of integration constants.
Note that eqs. \rf{levi} are Hamiltonian in the following sense
\[
e_1^\prime = \pder{{\cal H}_1}{c_1}=\left\{ e_1\, , \, {\cal
H}_1\right\}, \quad \quad 
c_1^\prime = -\pder{{\cal H}_1}{e_1}=\left\{ c_1\, , \, {\cal
H}_1\right\}\, ,
\]
where as follows from the previous subsection :
\begin{equation}
{\cal H}_1 = 2 x e_1 c_1-e_1^2 c_1-e_1 c_1^2 + {\bar k}_1 c_1 - k_1
e_1, \quad
\left\{ e_1\, , \,c_1\right\}=1\, .
\lab{newham1}
\end{equation}
Also we derive from 
\[ \pder{{\cal H}_1}{x}= 2 e_1 c_1
\]
that 
\[{\cal H}_{1\,xx}= 2  e_1^\prime c_1 + 2 e_1 c_1^\prime
=2 e_1^2 c_1 -2 c_1^2 e_1+2 {\bar k}_1 c_1+2 k_1 e_1
\]
and
\[
2 \left( {\cal H}_1 - x {\cal H}_{1\,x}\right)  =2 \left( - e_1^2 c_1 - c_1^2 e_1
+{\bar k}_1 c_1 - k_1 e_1 \right) 
\]
Accordingly,
\[
{\cal H}_{1\,xx} +2 \left( {\cal H}_1 - x {\cal H}_{1\,x}\right)  = 4 c_1 \left( 
e_1 c_1 - {\bar k}_1  \right) = 2 c_1 \left( {\cal H}_{1\,x}- 2 
{\bar k}_1  \right)
\]
and 
\[
{\cal H}_{1\,xx} -2 \left( {\cal H}_1 - x {\cal H}_{1\,x}\right)  = 
 4 e_1 \left( e_1 c_1 + k_1  \right)  = 2 e_1 \left( {\cal H}_{1\,x}+2 
k_1  \right)\, .
\]
Thus, ${\cal H}_1$ satisfies the Jimbo-Miwa equation \cite{jimbo} 
of  Painlev\'e IV system:
\[\left({\cal H}_{1\,xx} +2 \left( {\cal H}_1 - x {\cal H}_{1\,x}  \right)\right)
\left({\cal H}_{1\,xx} -2 \left( {\cal H}_1 - x {\cal H}_{1\,x}  \right)\right)
=
2 {\cal H}_{1\,x} \left( {\cal H}_{1\,x}- 2 
{\bar k}_1  \right)\left( {\cal H}_{1\,x}+2 
k_1  \right)
\]

Connection of $M=1$ example \rf{levi}
to $A^{(1)}_2$  symmetric Painlev\'e IV set
of equations
\[
\begin{split}
f_{0\, x}&= f_0 (f_1-f_2) + \alpha_0 \\
f_{1\, x}&= f_1 (f_2-f_0) + \alpha_1 \\
f_{2\, x}&= f_2 (f_0-f_1) + \alpha_2 
\end{split}
\]
with $\alpha_0 + \alpha_1 + \alpha_2 =-2$ can be made explicit by setting 
\[
f_i=-c_1,\;\;\; f_{i+1}=-e_1+\frac{c_{1\,x}}{c_1} \;\;\; f_{i+2}= e_1+c_1-\frac{c_{1\,x}}{c_1}-2x,
\;\; \alpha_i=k_1, \;\; \alpha_{i+2}=-k_1-{\bar k}_1
\]
for $i=0,1,2$ and
with the Darboux-B\"acklund transformation $g$ defined in \rf{g2bose}
and accordingly mapping $f_i$ to $f_{i+1}$, $\alpha_i \to - \alpha_{i+1}$ and $\alpha_{i+2} \to
\alpha_i+\alpha_{i+1}$.
This is consistent with realization of $g$ as $g =\pi s_{i}$ 
for $i=0,1,2$, where generators $s_i$ of the affine Weyl group $A^{(1)}_2$ act as
$s_i (\alpha_{i+2})=\alpha_i+\alpha_{i+2}$. 
The above solutions together with the idea of introducing 
permutation symmetry of the extended affine Weyl group $A^{(1)}_2$
by associating $f_i$'s to any of the solutions of the Levi system
was discussed in \cite{Aratyn:09}.

\subsection{The Four-Bose system and $\boldmath{ A^{(1)}_4}$
Painlev\'e Equations}
\label{subsection:M=2}
We now consider a four-boson case with  $M=2$
and $(c_k,e_k)_{k=1}^{2}$ subject to equations
\begin{equation}
\begin{split}
e_1^\prime&=2xe_1-(e_1+2c_1+2c_2)e_1+{\bar k}_1\\
e_2^\prime&=2xe_2-4xe_1-(e_2+2c_2)e_2+(2c_1+2e_1+
4c_2)e_1+{\bar k}_2\\
c_1^\prime &= -2xc_1+4xc_2+(c_1+2e_1)c_1+
(2c_1-2c_2-4e_2)c_2+k_1\\
c_2^\prime&=-2xc_2+(c_2+2e_2)c_2+k_2\\
\end{split}
\lab{4boseqs}
\end{equation}
as follows from  equations \rf{eprj}-\rf{eprj}.
The corresponding Hamiltonian is
\begin{equation}
\begin{split}
{\cal H}_2&= - e_{1} c_{1} \left( e_{1}+ c_{1}+2 c_2 \right)
- e_{2} c_2  \left( e_{2}+  c_2 \right)
+2 x \sum_{i=1}^2 e_ic_i \\
&-k_2 e_2 +{\bar k}_1 c_1 -(k_1+2 k_2)e_1 +({\bar k}_2 +2 {\bar k}_1)
c_2\, .
\lab{4boseham}
\end{split}
\end{equation}
{}We find that in the case of $M=2$ the vector fields $E_j,C_j,\,i=1,2$
are
\[
E_1= \pder{}{c_1}, \;\;\; E_2=\pder{}{c_2} -2 \pder{}{c_1},
\;\;\;C_1= 2\pder{}{e_2} - \pder{}{e_1} , \;\;\;
C_2= -\pder{}{e_2} 
\]
and according to \rf{pbec} they lead to the Poisson brackets  :
\begin{equation}
\left\{ e_1\, , \, c_1 \right\} = 1 , \;\;
\left\{ e_1\, , \, c_2 \right\} = 0 , \;\;
\left\{ e_2\, , \, c_1 \right\} = -2 , \;\;
\left\{ e_2\, , \, c_2 \right\} = 1
\lab{M2pb}
\end{equation}
consistent with equations of motion \rf{4boseqs} through
the Poisson brackets :
\begin{equation}
e_j^\prime=
\left\{ e_j \, ,\, {\cal H}_2\right\}, \qquad
c_j^\prime = \left\{ c_j \, ,\, {\cal H}_2\right\} , \quad j=1,2\, .
\lab{4bosepb}
\end{equation}

The symmetry transformations \rf{gemcm}-\rf{gaddition}
read here :
\begin{xalignat}{2}
g( e_2)&=e_{1} + c_2, &
g( e_{1})&=e_{1}-e_2 +c_2+c_{1} + \frac{c_2^{\prime}}{c_2}\lab{4bosong} \\
g( c_2)&=-e_{1} + e_2-\frac{c_2^{\prime}}{c_2},&
g( c_{1})&= e_{1} -\frac{\left(   -e_{1} + e_2-c_2-c_{1}
-\frac{c_2^{\prime}}{c_2} \right)^{\prime}}{\left(-e_{1} + e_2-c_2-c_{1}
-\frac{c_2^{\prime}}{c_2}\right)} \, ,\nonu
\end{xalignat}
which keep equations \rf{4boseqs} invariant for :
\begin{xalignat}{2}%\begin{equation} \begin{split}
g(k_1)&=-2+{\bar k}_1+2{\bar k}_2, &
g({\bar k}_1)&=2-{\bar k}_1-{\bar k}_2-k_1-3k_2, \nonu \\
g(k_2)&=2-{\bar k}_1-{\bar k}_2,& 
g({\bar k}_2)&=-4+3{\bar k}_1+2{\bar k}_2+2k_1+5k_2\,.
\lab{4bosongk}
\end{xalignat} %\end{split} \end{equation}

In order to see the meaning of this transformation from the 
group theoretic point of view we cast equations \rf{4boseqs} into the symmetric $A^{(1)}_4$ Painlev\'e equations:
\begin{equation} 
f_i^{\prime} = f_i (f_{i+1}- f_{i+2} + f_{i+3} - f_{i+4} ) + \alpha_i, \quad i=0,\ldots,4 
\lab{a14painleve}
\end{equation}
with conditions $f_i=f_{i+5}$ and $\sum_{i=0}^4\alpha_i=-2$. 
We propose the following identification
\[  \begin{split}
f_1 &=-e_1\\
f_2&=g (f_1)=- \left(e_{1}-e_2 +c_2+c_{1} + \frac{c_2^{\prime}}{c_2}\right)=
-e_1-e_2-c_1-2c_2+2x-\frac{k_2}{c_2} \\
f_3&=-c_2 \\
f_4&=g(f_3)= -\left(-e_{1} + e_2-\frac{c_2^{\prime}}{c_2}\right) =
e_{1} + e_2+c_2-2x+\frac{k_2}{c_2}\\
f_0&=-f_1 -f_2-f_3-f_4-2x=e_1+c_1+2c_2-2x
\end{split}
\]
and
\[  \begin{split}
\alpha_1 &=-{\bar k}_1,\qquad
\alpha_2=2-  {\bar k}_1- {\bar k}_2-k_1-3k_2   \\
\alpha_3&=k_2, \qquad
\alpha_4=-2 +{\bar k}_1  +{\bar k}_2\\
\alpha_0&=-2 +\sum_{i=1}^4 \alpha_i = -2 +{\bar k}_1+k_1+2k_2  \, .
\end{split}
\]
Alternatively we can write relations between  
$e_i, c_i,\,i=1,2$ and $f_i,\,i=0, 1,{\ldots},4$ as
\begin{equation}
\begin{split}
e_1&=-f_1,\;\quad e_2=-f_1 -f_4+f_3^\prime/f_3
=-f_0-f_2+\frac{\alpha_3}{f_3}\\
c_1&=f_3-f_2-f_4, \; \quad c_2=-f_3
\end{split}
\lab{ecf4bose}
\end{equation}
in agreement with equations \rf{eMM} and \rf{cMM}.
Accordingly one can rewrite the $g$-transformation from \rf{4bosong}
as
\begin{equation}
\begin{split}
g(f_1)&=f_2, \qquad g( f_3)=f_4\\
g( f_2)&= -f_3+f_2+f_4-\frac{f_4^{\prime}}{f_4}+
\frac{f_2^{\prime}}{f_2} \\
&=f_3-\frac{\alpha_4}{f_4}+\frac{\alpha_2}{f_2}\\
g( f_4)&= f_1+f_3-f_2+\frac{f_4^{\prime}}{f_4}
=f_0+\frac{\alpha_4}{f_4}\, .
\end{split}
\lab{4bosongf}
\end{equation}
Comparing with definitions of transformations $s_i, \, i=1,2,3,4$
(see f.i. \cite{noumibk}) we see that expression 
for transformation $g$ from \rf{4bosong}-\rf{4bosongk} agrees with
\[
g = \pi s_1+\pi s_3-\pi
\]
as applied on both $f$'s and $\alpha$'s.

More generally, associating $-e_1$ and $-c_2$ to $f_i$ and $f_{i+2}$, respectively, with
$i$ taking all the values $i=1,2,3,4$ we obtain identifications:
\[
g = \pi s_i+\pi s_{i+2}-\pi \equiv g_i
\]
for each realization, where
$s_i$ are  generators of the affine Weyl $A^{(1)}_4$. 
This relates $g$ with $s_i+ s_{i+2}$ and by varying $i$ over all its
values makes possible to recover all the affine Weyl $A^{(1)}_4$ generators $s_i$ from one 
Darboux-B\"acklund transformation $g$.
For instance $s_1=(-I+\pi(g_1+g_2+g_3-g_0-g_4))/2$.

\subsection{$\boldmath{M=3}$ Bose System and the symmetric $\boldmath{A^{(1)}_6}$  Painlev\'e Equations}
For $M=3$,  equations \rf{eprj}-\rf{cprj} become
\begin{equation}
\begin{split}
e_1^\prime&=2xe_1-(e_1+2c_1+2c_2+2c_3)e_1+{\bar k}_1\\
e_2^\prime&=2xe_2-4xe_1-(e_2+2c_2+2c_3)e_2-(-2c_1-2e_1-
4c_2-4c_3)e_1+{\bar k}_2\\
e_3^\prime&=2xe_3-4xe_2+4xe_1-(e_3+2c_3)e_3-(-2c_2-2e_2-4c_3)e_2
-(2c_1+2e_1+4c_2+4c_3)e_1+{\bar k}_3\\
c_1^\prime &= -2xc_1+4xc_2-4xc_3+(c_1+2e_1)c_1+
(2c_1-2c_2-4e_2)c_2+(2c_1-4c_2+2c_3+4e_3)c_3+k_1\\
c_2^\prime&=-2xc_2+4xc_3+(c_2+2e_2)c_2+(2c_2-2c_3-4e_3)c_3+k_2\\
c_3^\prime&=-2xc_3+(c_3+2e_3)c_3+k_3\, .
\lab{M3-eqs}
\end{split}
\end{equation}
These equations are Hamiltonian as in eqs.\rf{ejcjHeqs} 
with
\begin{equation}
\begin{split}
{\cal H}_3 &= 2\,x\,{e_1}\,{c_1} + 2\,x\,{e_2}
\,{c_2} + 2\,x\,{e_3}\,{c_3} - {e_1}\,
{c_1}\,({e_1} + {c_1} + 2\,{c_2} + 2\,
{c_3}) \\
&- {e_2}\,{c_2}\,({e_2} + {c_2} + 
2\,{c_3}) - {e_3}\,{c_3}\,({e_3} + 
{c_3}) + {{\bar k}_1}\,{c_1} + ({{\bar k}_2} + 2\,
{{\bar k}_1})\,{c_2} \\
&+ ({{\bar k}_3} + 2\,{{\bar k}_1} + 2\,{{\bar k}_2})\,
{c_3} - ({k_1} + 2\,{k_3} + 2\,{k_2})\,
{e_1} - ({k_2} + 2\,{k_3})\,{e_2} - 
{k_3}\,{e_3}\, . 
\end{split}
\lab{6boseham}
\end{equation}
The symmetric $A^{(1)}_6$  Painlev\'e equations
\begin{equation} 
f_i^{\prime} = f_i (f_{i+1}- f_{i+2} +f_{i+3}- f_{i+4}+f_{i+5} - f_{i+6} ) + \alpha_i, \quad
i=0,1,2, \ldots, 6, 
\lab{a16painleve}
\end{equation}
with condition $f_i=f_{i+7}$ and
with
\[
f_0= -\sum_{i=1}^6 f_i -2 x
\]
are satisfied by 
\begin{equation} \begin{split}
{f_0} &= {e_1} + {e_2} + {c_2} + 2\,
{c_3} - 2\,x \\
{f_1} &= {e_1} + {c_1} + 2\,{c_2} + 2\,
{c_3} - 2\,x\\
{f_2} &=  - {e_1}\\
{f_3} &=  - {e_2} - {e_1} - {c_1} - 2\,
{c_2} - 2\,{c_3} + 2\,x\\
{f_4} &= -e_2+e_3-c_2-c_3-\frac{c_3^{\prime}}{c_3}\\
&= - {e_2} - {e_3} - {c_2} - 2\,
{c_3} + 2\,x - {\displaystyle \frac {{k_3}}{{
c_3}}} \\
{f_5} &=  - {c_3}\\
{f_6} &= -\left( e_3-e_2-\frac{c_3^{\prime}}{c_3}\right)\\
&={e_2} + {e_3} + {c_3} - 2\,x + 
{\displaystyle \frac {{k_3}}{{c_3}}} 
\lab{fi-A16}
\end{split}
\end{equation}

Furthermore
\[ \begin{split}
{\alpha_0} &=  - 2 + {k_2} + 2\,{k_3} + {
{\bar k}_2} + {{\bar k}_1} =-2 +\kappa_2+{\bar \kappa}_2- {\bar
\kappa}_1\\
{\alpha_1} &=  - 2 + {k_1} + 2\,{k_2} + 2\,
{k_3} + {{\bar k}_1}=-2+\kappa_1+{\bar \kappa}_1\\
{\alpha_2} &=  - {{\bar k}_1}=-{\bar \kappa}_1\\
{\alpha_3} &= 2 - {k_1} - 2\,{k_2} - 2\,{k_3
} - {{\bar k}_2} - {{\bar k}_1}=2-\kappa_1- {\bar \kappa}_2+ 
{\bar\kappa}_1\\
{\alpha_4} &= 2 - {{\bar k}_2} - {{\bar k}_3} - {k_2} - 
3\,{k_3}= 2 -{\bar \kappa}_3 +{\bar \kappa}_2 -\kappa_2-\kappa_3\\
\alpha_5 &= k_3=\kappa_3\\
{\alpha_6} &=  - 2 + {{\bar k}_2} + {{\bar k}_3}=-2 
+{\bar \kappa}_3 -{\bar \kappa}_2
\end{split}
\]
in terms of objects  $e_i,c_i,k_i, {\bar k}_i, \; i=1,2,3$ from $M=3$ equations \rf{M3-eqs}.

The logic of deriving the above association is as follows.
We initially set $f_5=f_{2M-1} = -c_M=-c_3$
and $f_6=f_{2M}= g (f_5)$. This is suggested by equation of motion for $c_M$ 
which has the unique $ 2x c_M$  term
on the right hand side making it a good candidate for $f_i $, since Painlev\'e equations have this structure after elimination of $f_0$.
But equations of motion for $f_5$ and $f_6$ after elimination of $f_0$ only involve sums  $f_1+f_3$
and $f_2+f_4$, respectively. Therefore these quantities are derived from these equations to be $f_1+f_3=-e_2=-e_{M-1}$ and $f_2+f_4= g(f_1+f_3)=-g(e_2)$.
Next, we turn into equation for $f_1$  in \rf{a16painleve}, which we rewrite as
\[
f_1^{\prime}= f_1 \left(-2e_1-2c_3-2c_2+2x+f_1 \right) +\alpha_1
\]
after we substituted $f_5=-c_3, f_1+f_3=-e_2$ and 
$f_0=-(f_2+f_4)-e_3-k_3/c_3$ and the values for 
$f_6$ and $f_2+f_4$ determined previously by $g$ transformations.
We find that the solution to these equations is given by
\[ \begin{split}
{f_1} &= {e_1} + {c_1} +2 {c_2} + 2\,
{c_3} - 2\,x \\
{\alpha_1} &=  2 + {k_1} + 2\,{k_2} + 2{
{\bar k}_3} + {{\bar k}_1}
\end{split}
\]
{}From that result we derive $f_3$ as $-e_2-f_1$ and from eq. for $f_3$ we derived $f_4$ and $f_2=-e_1$.

Applying similarity transformation \rf{simM}
\[
L_3 \to (\partial-e_3)^{-1} L_3(\partial-e_3)
\]
on the Lax operator obtained from \rf{1c} by setting $M=3$. 
 \begin{equation}
\begin{split}
g( e_3)&=e_{2} + c_3, \qquad g( c_3)=-e_{2} + e_3-\frac{c_3^{\prime}}{c_3}\\
g( e_{2})&=e_{1}+e_{2}-e_3 +c_3+c_{2} + \frac{c_3^{\prime}}{c_3} \\
g( c_{2})&=-e_{1} + e_{2} -\frac{\left(   -e_{2} + e_3-c_3-c_{2} -\frac{c_3^{\prime}}{c_3} \right)^{\prime}}{\left(-e_{2} + e_3-c_3-c_{2} -\frac{c_3^{\prime}}{c_3}\right)}
\end{split}
\lab{gecM6}
\end{equation}
in addition to
\[
g \left( e_1+c_2 +c_3 \right)=\sum_{l=1}^3 c_l\,.
\]

These B\"acklund transformations amount to
the following transformations when applied on $f_i$ : 
\begin{xalignat}{2}
g(f_2) &= f_3-\frac{\alpha_4}{f_4}, &
g (f_1+f_3)&=f_2+f_4\nonu\\
g(f_0)&= f_1, &
g(f_4)&=f_5+\frac{\alpha_4}{f_4} -\frac{\alpha_6}{f_6}\nonu\\ 
g(f_5)&=f_6, &
g(f_6)&=f_0+\frac{\alpha_6}{f_6}, \nonu
\end{xalignat}
which agrees with the action of
\begin{equation}
g = \pi s_5+\pi s_3 -\pi
\lab{gs5s3}
\end{equation}
as applied on $f_i$ and $\alpha_i$.

Now the B\"acklund transformation of $c_1$ is equal to 
\begin{equation}
g\left( c_1\right) = e_1-\frac{\left(g(c_2)+g(c_3) -c_1-c_2-c_3\right)^{\prime}}{g(c_2)+g(c_3) -c_1-c_2-c_3}
\lab{gcone}
\end{equation}
in terms of $g(c_2)$ and $g(c_3)$ from eqs. \rf{gecM6}.

In terms of $f_i$ and $\alpha_i$'s it takes a form
\[
\begin{split}
g\left( c_1\right) &= -f_2 - \frac{\left(f_3-\frac{\alpha_4}{f_4}\right)^{\prime}}{f_3-\frac{\alpha_4}{f_4}}\\
&= -f_1-f_4-f_6+f_0+f_5+\frac{\alpha_4}{f_4}
-\frac{\alpha_3+\alpha_4}{f_3-\frac{\alpha_4}{f_4}}
\end{split}
\]
This gives rise to the following transformation of $f_1$:
\[
g(f_1) = f_2 - \frac{\alpha_3+\alpha_4}{f_3-\frac{\alpha_4}{f_4}}
\]
Thus representation of the B\"acklund transformation $g$ in terms 
of generators of the extended affine Weyl group $A^{(1)}_6$
from \rf{gs5s3} needs to be augmented as follows:
\[
g = \pi s_5+\pi s_3 -\pi + s_4 \left(\pi s_2 -\pi\right)=
 \pi s_5+\pi s_3 s_2 -\pi 
\]
as applied on $f_i$ and $\alpha_i$.

By generalizing eq. \rf{fi-A16} by substituting $i=5$ with arbitrary
$i$ between $0$ and $M$ one obtains
\[
\begin{split}
g_i &= \pi s_i+\pi s_{i-2} s_{i-3}  -\pi,\;\;\;  i=0,1, \ldots, 6\\
f_i &= -c_M
\end{split}
\]
providing a scheme to reproduce all generators 
$s_0, s_1, \ldots, s_6$ of the affine Weyl group of $A^{(1)}_6$
by one B\"acklund transformation $g$.

\section{Reduction of $\boldmath{M=2}$ case. Painlev\'e V}
\label{section:reduction}
We will follow reference \cite{Aratyn:1994vc} and perform
a Dirac reduction of $M=2$ case (see subsection \ref{subsection:M=2}),
by redefining  variables as follows :
\[
\left( e_1, c_1, e_2 ,c_2 \right) \rightarrow 
\left( {\tilde e}_1=e_1, {\tilde c}_1=c_1, {\tilde e}_2=(e_2-c_2)/2 ,
{\tilde c}_2=(c_2-e_2)/2 \right) \, ,
\]
which is equivalent to setting a second-class constraint 
\[
c= c_2=-e_2
\]
with the Dirac bracket:
\[ \left\{ c , c \right\}= \frac12 \delta_x (x-y)\, .
\]
The self-similarity reduction applied on the resulting
$t_2$ evolution equations \rf{t2flow} yields
\begin{align}
-2 x c_1&= c^{\prime}_{1}+2 c^{\prime}-c^{2}_{1}-2 e_{1} c_{1}-2 c_{1}\, c+ 
k_1 \lab{eqc1} \\
-2 x e_1 &=  -e^{\prime}_{1}-e^{2}_{1}-2 e_{1} c_{1}-2 e_{1} \,c  +{\bar k}_1 
\lab{eqe1} \\
-2 x c&=  e^{\prime}_{1}+ e_{1} c_{1} +k \, . \lab{eqc}
\end{align}
Eliminating $c$ and $c_1$ from equations \rf{eqc1}-
\rf{eqc} yields the following expression for
$y=e_1/2x$ :
\begin{equation} \begin{split}
y_{zz}& = - \frac1z y_z+ \left( \frac{1}{2y}+\frac{1}{2(y-1)} \right) y_z^2
- \frac{\alpha y}{z^2 (y-1)} \\
&-\frac{\beta (y-1)}{z^2 y} - \frac{\gamma}{z} y (y-1) 
-\delta y (y-1) (2y-1)
\lab{P5y}
\end{split}
\end{equation}
with constants
\begin{equation}
\begin{split}
\alpha&= \frac18 (k+1)(k+{\bar k}_1+1)+ \frac{{\bar k}_1^2}{32}
=\frac18\left(k+1+\frac{{\bar k}_1}{2}\right)^2 \\
\beta&=-\frac{{\bar k}_1^2}{32}= -\frac18\left(\frac{{\bar
k}_1}{2}\right)^2 , \quad
\gamma= \frac{k+ k_1+1}{2 \sigma}, \quad \delta =-\frac{1}{2 \sigma^2}
\lab{abc}
\end{split}
\end{equation}
after a change of coordinate $x \to z$ such that $z= \sigma x^2$.

The above equation takes on a conventional form of the Painlev\'e V
equation for $w=y/(y-1)$ and $\delta=-1/2$.

To study the Darboux-B\"acklund transformation of  the Painlev\'e V 
system we perform the similarity transformation
\[
L \to \left(\partial+c\right) ^{-1} L \left(\partial+c\right) 
\]
on the Lax operator for the reduced 4-boson system \cite{Aratyn:1994vc}.
This induces the following transformations for variables of the reduced subspace:
\[ \begin{split}
g\left( e_1 \right)&=e_1+c_1+2c \\
g\left( c_1 \right)&=e_1-\frac{\left(c_1+e_1+2c\right)^{\prime}}{c_1+e_1+2c}\\
g \left( c\right)&=-e_1-c 
\end{split}
\]
It follows that $g$ transforms the constants
$k,k_1, {\bar k}_1$ as
\begin{equation} \begin{split}
g(k) &= -k-{\bar k}_1\\
g({\bar k}_1) &= k_1+{\bar k}_1+2k\\
g(k_1) &= {\bar k}_1-2
\lab{gks}
\end{split}
\end{equation}
Next, applying this transformation to solution $w=y/(y-1) $ 
of a  conventional Painlev\'e V equation
yields
\begin{equation}
\begin{split}
g (w) &=1 - \frac{2 z \sigma w }{F} \\
F&= + z w_z - \frac12 w^2 \left(k+1+\frac{{\bar k}_1}{2}\right)
+ w \left( \frac12 (k+1)+z \sigma \right) + \frac14 {\bar k}_1
\lab{gw}
\end{split}
\end{equation}
In terms of quantities
\[
c_g= \frac12  \left(k+1+\frac{{\bar k}_1}{2}\right), \;\;
a_g = \frac14 {\bar k}_1
\]
with properties
\[c_g^2 = 2 \alpha , \;\;\; a_g^2 = - 2 \beta
\]
the function $F$ from relation \rf{gw} can be rewritten as
\[
F= + z w_z -  w^2 c_g
+ w \left( c_g-a_g+z \sigma \right) + a_g
\]
in complete agreement with \cite{Gromak1976,Gromak2002}.

\section{Outlook}
\label{section:outlook}
We have here derived the higher order Painlev\'e equations by taking self-similarity 
limit of the special class of integrable models and shown 
how the extended affine Weyl groups $A^{(1)}_n$ symmetries are induced
by B\"acklund transformations generated by translations on the underlying
``half-integer'' Volterra lattice. 
The Hamiltonian of the integrable model reduced by the self-similarity
procedure
has been explicitly shown to transform under change of variables into
the Hamiltonian for the higher Painlev\'e equations.
In the forthcoming publication we plan to provide explicit proof for 
formulas governing such change of variables and include in the
formalism the Painlev\'e equations with the extended affine Weyl 
groups $A^{(1)}_{2n-1}$ symmetries.
We will also employ a link between
on the one hand integrable hierarchies and on the other hand higher 
order Painlev\'e equations to derive  the corresponding
higher order Painlev\'e hierarchies.

\vskip .4cm \noindent
{\bf Acknowledgments} \\
JFG and AHZ thank CNPq and FAPESP  for partial
financial support.
Work of HA was partially supported by FAPESP.
HA thanks Nick Spizzirri for discussions.
The authors thank Danilo Virges Ruy for discussions.

\end{document}